\documentstyle[preprint,aps]{revtex}

\begin{document}
\draft
\preprint{Dortmund, May 1995}

\title{A Microscopic Model of Non-Reciprocal Optical Effects in Cr$_2$O$_3$}
\author{V. N. Muthukumar, Roser Valent\'\i\ and Claudius Gros
       }
\address{Institut f\"ur Physik, Universit\"at Dortmund,
         44221 Dortmund, Germany
        }
\date{\today}
\maketitle
\begin{abstract}
We develop a microscopic model that explains non-reciprocal optical effects
in centrosymmetric Cr$_2$O$_3$. It is shown that light can couple
{\em directly} to the antiferromagnetic order parameter. This coupling
is mediated by the spin-orbit interaction and involves an
interplay between the breaking of inversion symmetry due to the
antiferromagnetic order parameter and the trigonal field contribution
to the ligand field at the Cr$^{3+}$ ion.
\end{abstract}
The study of the interaction of light with magnetic substances has a long
history. A classic example is the Faraday effect in ferromagnets where
light couples directly to the ferromagnetic order parameter.  Since the
pioneering work of Argyres \cite{Argyres}, it is known that
electromagnetic radiation couples to the internal molecular field in a
ferromagnet (which in turn is proportional to the order parameter)
through the spin-orbit interaction.  Such a coupling would of course be
absent in antiferromagnets, where the internal molecular field is zero.
In the absence of such a direct coupling between light and the
antiferromagnetic order parameter, antiferromagnetic ordering could so
far be probed only {\it indirectly}, for instance, by Raman scattering
of magnetic excitations \cite{Raman}.

The discovery \cite{Krichevtsov_93,Fiebig_94} of non-reciprocal optical
effects (i.e., not invariant under time reversal) \cite{S_P_92} below
the N\'eel temperature, $T_N$, in optical experiments on Cr$_2$O$_3$ has
therefore been considered a breakthrough in the study of antiferromagnetic
ordering by light for it is only this class of experiments which can
distinguish between two magnetic states that are related to each other by the
time-reversal operation. Fiebig {\it et al.} \cite{Fiebig_94} have found that
antiferromagnetic domains could be observed {\it directly} by non-reciprocal
second harmonic generation (SHG), leading to the first photographs ever
of antiferromagnetic domains \cite{APL,note_Roth}. These experiments
show that light can indeed couple directly to the antiferromagnetic
order parameter. Though such a coupling was anticipated earlier from
symmetry considerations \cite{Brown}, no microscopic mechanism has been
presented so far.

In this Letter, we present a microscopic mechanism that explains all
non-reciprocal optical effects in Cr$_2$O$_3$.  While the spin-orbit
interaction is, of course, essential in coupling the charge with the spin
degrees of freedom it does not suffice for the generation of
non-reciprocal effects. We find that non-reciprocal effects arise from an
interplay between the breaking of crystal inversion symmetry by the
antiferromagnetic order parameter and the trigonal distortion of the
ligand field at the Cr$^{3+}$ ion. This effect, in addition to the
spin-orbit interaction,
leads to a coupling of the antiferromagnetic order parameter with light.

We present furthermore a simple cluster model, containing the full crystal
symmetry of Cr$_2$O$_3$, which allows, for the first time, the
orders of magnitude of all matrix elements contributing
to the non-reciprocal phenomena in Cr$_2$O$_3$ to be predicted.
We apply the microscopic model to the observed phenomenon of SHG
\cite{Fiebig_94} and explain how antiferromagnetic domains can be
distinguished experimentally. We also apply our model to another
non-reciprocal effect seen experimentally in Cr$_2$O$_3$ viz.,
gyrotropic birefringence \cite{Krichevtsov_93} and solve the long-standing
question regarding its order of magnitude.

As an introduction to SHG in Cr$_2$O$_3$, we discuss
the macroscopic theory \cite{Fiebig_94} in brief.
Above $T_N$ ($\simeq$ 307K), Cr$_2$O$_3$ crystallizes
in the centrosymmetric point group $D_{3d}$.
The four Cr$^{3+}$ ions in the unit cell occupy equivalent $c$ positions
along the $C_3$ (optic) axis. Since this structure has a center
of inversion, parity considerations allow
only magnetic dipole transitions related to the existence of
an axial tensor of odd rank. Below $T_N$, time reversal
symmetry (R) is broken and in this case, since Neumann's principle
\cite{Birss} cannot be applied to non-static phenomena, only symmetry
operations of the crystal that do not include R may be used to classify
the allowed tensors for the susceptibilities. For Cr$_2$O$_3$, the
remaining invariant subgroup is $D_3$. New tensors are allowed in this
point group, for instance, a polar tensor of odd rank, that allow
electric dipole transitions in SHG.

{}From Maxwell's equations, one can derive the expression for SHG
by considering the contributions to (non-linear)
magnetization ${\bf M}^{(2 \omega)}=
{\bf{\gamma}}^{(2 \omega)} : {\bf E}^{(\omega)}~ {\bf E}^{(\omega)}$ and
polarization
${\bf P}^{(2 \omega)}= {\bf \chi}^{(2 \omega)} :{\bf E}^{(\omega)}~
{\bf E}^{(\omega)}$. The source term ${\bf S}({\bf r},t)$ in the wave equation
$ [ {\bf{\nabla}} \times ( {\bf{\nabla}} \times ) + \left( 1/c^2\right)
\partial^2/\partial t^2 ] {\bf{E}}( {\bf r}, t) =
-{\bf{S}}({\bf r}, t)$ can be written in a dipole expansion as \cite{Rosen},
\begin{equation}
{\bf{S}}({\bf r},t) =
\mu_o~ \left( \frac{{\partial}^2 {\bf{P}}({\bf r},t)}{\partial t^2} +
{\bf{\nabla}} \times \frac{\partial {\bf{M}}({\bf r},t)}{\partial t}+...
\right)~~.
\label{source}
\end{equation}
Then, by assuming that {\bf E}, {\bf P} and {\bf M} can be decomposed
in a set of plane waves and considering a circular basis with
${\bf{E}} = E_+ {\bf{e}}_++ E_- {\bf{e}}_- + E_z {\bf{e}}_z $, where
${\bf{e}}_{\pm} = \mp\frac{1}{\sqrt{2}} ({\bf{e}}_x {\pm} i {\bf{e}}_y)$
and the direction of laser light to be along the optic axis,
one obtains \cite{Fiebig_94},
\begin{equation}
{\bf{S}} = \left( \begin{array}{c}
S_+ \\ S_- \\ S_z \end{array} \right) = \frac{4 \sqrt{2} {\omega}^2} {c^2}
\left(\begin{array}{c}
(-\gamma_m + i \chi_e) E_-^2 \\  (\gamma_m + i \chi_e) E_+^2  \\ 0
\end{array} \right) ~~,
\label{S}
\end{equation}
where $\gamma_m$ and $\chi_e$ are non-zero components of the magnetic
($\gamma$) and
electric  ($\chi$) susceptibilities that are allowed by $D_3$ symmetry.
Incoming right circularly polarized light ($E_+$) leads to left circularly
polarized light ($E_-$) and vice versa in SHG.
Above $T_N$, as $\chi_e \equiv 0$, the SHG intensities $I_{\pm}$ are
identical while below $T_N$, the intensities
$I_{\pm} \propto | \pm\gamma_m + i \chi_e |^2 E_{\mp}^4$
are different for right and left
circularly polarized light, as observed in experiment
\cite{Fiebig_94}.

The macroscopic theory dicussed so far is based on symmetry
considerations only and does not provide estimates of the
magnitude of either $\gamma_m$ or $\chi_e$, which can be
obtained only from a microscopic approach. A {\em two ion}
mechanism for a non-zero electric dipole matrix element, $\chi_e$,
in non-centrosymmetric antiferromagnets was proposed
by Tanabe {\it et al.} \cite{Tanabe_65}.
This mechanism is unlikely to be responsible for
the coherent interference effect suggested by macroscopic theory (\ref{S}),
since it involves magnetic excitations.
Here, we propose instead, a {\em single ion}
mechanism. In our theory the coupling of light
to the antiferromagnetic order parameter is
through the interference between the coherent
contributions from the distinct Cr$^{3+}$ ions in
the crystal unit cell. No magnetic excitations are
involved in our coupling mechanism.

The crystal field due to the oxygen ions in Cr$_2$O$_3$ splits the
five-fold degenerate $3d$-orbitals of the Cr$^{3+}$ ions into two
levels, the lower one ($t_2$ level) being triply degenerate ($d_{xy}$,
$d_{yz}$, $d_{zx}$) and the upper ($e$ level) doubly degenerate
($d_{x^2-y^2}$, $d_{3z^2-r^2}$). The three $t_2$ orbitals are occupied in
the ground state and the two $e$ orbitals are empty. In SHG, the
Cr$^{3+}$ ion absorbs two photons and is excited to a $(t_2)^2e$
configuration via two consecutive
electric dipole processes (ED), corresponding to an
${\bf r}\cdot{\bf E}^{(\omega)}$ term in the Hamiltonian.
A contribution to $\gamma_m$ is then obtained for an
emission via a magnetic dipole process (MD), corresponding to an
${\bf L}\cdot{\bf B}^{(2\omega)}$ term in the Hamiltonian.
This contribution to $\gamma_m$ is allowed at all temperatures.
The key point of a microscopic model is then to find the
mechanism which allows a contribution to $\chi_e$ via
an electric dipole matrix element,
${\bf r}\cdot{\bf E}^{(2\omega)}$, in emission.

We present our theory in two steps. As a first step,
we explain the coupling mechanism between
light and the order parameter and later,
we proceed to discuss the role of the D$_{3d}$ crystal symmetry.
In order to understand the origin of the ED transition below $T_N$,
let us consider, for the moment, just
two $d$-orbitals per Cr$^{3+}$ ion:
the $d_{xy}$ orbital (ground state) and the $d_{x^2-y^2}$ orbital (excited
state). The MD contribution to SHG, i.e., to ${\bf M}^{(2\omega)}$ is
$$
\langle d_{xy},m_s| {\bf L} |d_{x^2-y^2},m_s\rangle
\langle d_{x^2-y^2},m_s| ({\bf r}\cdot{\bf E^{(\omega)}})^2|d_{xy},m_s
\rangle ~,
$$
where the label $m_s$ denotes the spin quantum number of the relevant
orbital. (Energy denominators have been omitted for clarity.) Here
$\ ({\bf r}\cdot{\bf E^{(\omega)}})^2\ $ denotes
$\ {\bf r}\cdot{\bf E^{(\omega)}}|\alpha \rangle
\langle \alpha|{\bf r}\cdot{\bf E^{(\omega)}}\ $
together with a sum over the intermediate states $|\alpha \rangle$ and
corresponding energy denominators.

In order to develop the theory for the ED contribution, we treat the
spin-orbit interaction and the trigonal distortion of the ligand field
at the Cr$^{3+}$ ion
as perturbations to the $3d$ Cr states. It is easy to see
that the diagonal part of the spin-orbit interaction mixes the $t_2$ and
$e$ states, viz.,
\begin{equation}
\langle d_{xy},m_s| {\bf L}\cdot{\bf S} |d_{x^2-y^2},m_s\rangle~ =~ 2i
\langle S_z \rangle ~,
\label{LS_ME}
\end{equation}
where $\langle S_z\rangle \equiv \langle m_s | S_z |m_s \rangle$.

Next, we note that the $3d$ and the $4p$ Cr orbitals are mixed by the
trigonal field \cite{trigonal}, which breaks local parity.
The local eigenstates of the Cr$^{3+}$ ion can therefore be written in the
lowest order of perturbation theory as
\begin{eqnarray}
|\tilde d_{xy}\rangle  & = & |d_{xy},m_s\rangle
                   + \lambda \langle S_z\rangle |d_{x^2-y^2},m_s\rangle
                   + \eta^{\prime}|p^{\prime},m_s\rangle
\nonumber\\
|\tilde d_{x^2-y^2}\rangle  & = & |d_{x^2-y^2},m_s\rangle
                   - \lambda \langle S_z \rangle |d_{xy},m_s \rangle
                   + \eta|p,m_s\rangle~~,
\nonumber
\end{eqnarray}
where $\lambda$ is proportional to the spin-orbit
coupling; $\eta$, $\eta^{\prime}$ are proportional to the
trigonal field and $|p,m_s\rangle, |p^{\prime},m_s\rangle$ are Cr $4p$
orbitals that will be specified later. The ED contribution to the SHG,
i.e., to
${\bf P}^{(2\omega)}$, can now be expanded in powers of $\lambda$ and $\eta$
as
\begin{eqnarray}
\langle \tilde d_{xy}| {\bf r}|\tilde d_{x^2-y^2}\rangle
\langle \tilde d_{x^2-y^2}| ({\bf r}\cdot{\bf E^{(\omega)}})^2|
\tilde d_{xy}\rangle  ~=
\nonumber \\
\eta\lambda \langle d_{xy},m_s| {\bf L}\cdot{\bf S} |d_{x^2-y^2},m_s\rangle
\langle d_{x^2-y^2},m_s| {\bf r} |p,m_s\rangle
\label{result} \\
\cdot \langle d_{x^2-y^2},m_s|( {\bf r}\cdot{\bf E^{(\omega)}})^2|d_{xy}
,m_s\rangle + \dots\ ,
\nonumber
\end{eqnarray}
where the contribution $\sim \lambda \eta $ is shown for illustration.
All non zero-contributions are $\sim\lambda\eta \langle S_z \rangle $.
To see this, consider all possible contributions order by order.

The contribution $\sim\lambda^0\eta^0$ vanishes because
all $d$-orbitals have even parity, viz.,
$\langle d_{xy},m_s| {\bf r} |d_{x^2-y^2},m_s\rangle \equiv 0$.
The contribution $\sim\lambda^0\eta^1$, viz.,
\begin{equation}
\eta \langle d_{xy},m_s| {\bf r} |p,m_s \rangle
     \langle d_{x^2-y^2},m_s|({\bf r}\cdot{\bf E}^{(\omega)})^2 |d_{xy},m_s
     \rangle
\label{non_zero}
\end{equation}
is finite for every Cr site and is proportional to $\eta$. However, note
that the trigonal field at Cr sites that are related by inversion
symmetry have opposite signs, {\it viz.}, if
A1/B2 and B1/A2 are pairs of Cr$^{3+}$ ions in the unit cell that are
related by inversion symmetry, $\eta_{A_1} = -\eta_{B_2}$ and
$\eta_{A_2} = -\eta_{B_1}$ \cite{trigonal}. On summing the
contributions to the matrix element $\sim \lambda^0\eta^1$ from each Cr
ion in the unit cell, we see that this matrix element has to vanish
identically.

The term $\sim\lambda^1\eta^0$ vanishes, again because of the
parity of the $d$-orbitals but the contribution $\sim\lambda^1\eta^1$
shown in (\ref{result}) does not vanish generally.  Using Eq.\ (\ref{LS_ME}),
we can sum up the total contribution $\sim\lambda^1\eta^1$ from the
four (equivalent) Cr ions in the unit cell of Cr$_2$O$_3$ as
\begin{eqnarray}
\lambda \eta {\bf X}^{(2\omega)} \,
     (\langle S_z \rangle _{A1}-\langle S_z\rangle _{B1}+\langle S_z\rangle
     _{A2}
\nonumber \\
\qquad-\langle S_z\rangle _{B2})=\lambda \eta {\bf X}^{(2\omega)}
\ \triangle(T)~,
\nonumber
\end{eqnarray}
where
${\bf X}^{(2\omega)}=\langle d_{x^2-y^2},m_s| {\bf r} |p,m_s \rangle
\langle d_{x^2-y^2},m_s|( {\bf r}\cdot{\bf E}^{(\omega)})^2|d_{xy},m_s
\rangle $ at site A1.
Thus, we have shown that the ED contribution to the SHG,
${\bf P}^{(2\omega)}$, couples directly to the
spin part of the antiferromagnetic order parameter
$\triangle (T) = \langle S_z\rangle _{A1}-\langle S_z\rangle _{B1}+
\langle S_z\rangle _{A2}-\langle S_z\rangle _{B2}$.
Note that ${\bf P}^{(2\omega)}$ is non-reciprocal, as it changes sign under
time-reversal and allows us to differentiate
between the different antiferromagnetic domains in Cr$_2$O$_3$.

We now complete the microscopic description by incorporating the full
symmetry of Cr$_2$O$_3$, i.e., $D_{3d}$.  This is done by considering a
(CrO$_6$)$_2$ cluster model shown in Fig.\ \ref{model}. In order to
reproduce the full D$_{3d}$ symmetry, one has to choose the locations
of the oxygen ions in this cluster in such a way that they do not
coincide exactly with those in the actual crystal structure.
This is because the cluster
model contains only two Cr sites, while there are four in the unit cell
of Cr$_2$O$_3$.

We define the Cr sites to be located at $z_o(0,0,\pm1)$ in the cluster
model ($z_o$ is an arbitrary constant).
 Neglecting the trigonal distortion for the moment, let us choose
the oxygens around the Cr sites to be located at
\begin{eqnarray}
\nonumber
z_o(0,0,\epsilon)&+& \vartheta\epsilon(\sqrt{2/3},0,\sqrt{1/3})\\
\label{sites_model}
z_o(0,0,\epsilon)&+& \vartheta\epsilon
(-\sqrt{1/6},\sqrt{1/2},\sqrt{1/3})\\
z_o(0,0,\epsilon)&+& \vartheta\epsilon
(-\sqrt{1/6},-\sqrt{1/2},\sqrt{1/3})~,
\nonumber
\end{eqnarray}
where $\vartheta=\pm1$ in order to obtain the sites of
all the 12 oxygen ions. (see Fig.\ \ref{model}). For any of the
four combinations $(\epsilon=\pm1, \vartheta=\pm1)$, the three
oxygen sites given in (\ref{sites_model}) form a plane. It is obvious
that this cluster model has a center of inversion at $(0,0,0)$. In
addition, it is easy to verify that this model has all the symmetry
elements belonging to the group $D_{3d}$ \cite{Birss}, in particular,
the $C_3$ and the $2_y$ symmetries that correspond to three- and
two-fold rotations about the $z$- and $y$-axes respectively.

The crystal field at the Cr site in the (CrO$_6$)$_2$
cluster is, in first approximation, cubic. The trigonal distortion of
the ligand field can be treated as a perturbation.  On doing this and
rotating the crystal field axes with respect to the crystallographic
axes (to facilitate calculations), we find that a convenient
representation for the Cr one-particle orbitals is given by
\begin{equation}
\begin{array}{rclcl}
t_2^{(1)} \ &=&\ -|d_{3z^2-r^2}\rangle  &+& \eta_1\,
                  |p_{z}\rangle\\
t_2^{(2)} \ &=&\ 3^{-1/2} [ \sqrt 2 |d_{x^2-y^2}\rangle
                      +  |d_{zx}\rangle] &+& \eta_2\,
		         |p_x\rangle  \\
t_2^{(3)} \ &=&\ 3^{-1/2} [ \sqrt 2 |d_{xy}\rangle
                      +  |d_{yz}\rangle ] &+& \eta_2\,
		         |p_y\rangle \\
  e^{(1)} \ &=&\ 3^{-1/2} [ |d_{xy}\rangle
                      - \sqrt 2 |d_{yz}\rangle ] &+& \eta_3\,
		        |p_y\rangle  \\
  e^{(2)} \ &=&\ 3^{-1/2} [ |d_{x^2-y^2}\rangle
                      - \sqrt 2 |d_{zx}\rangle] &+& \eta_3\,
		        |p_x\rangle~~,
\end{array}
\label{states}
\end{equation}
where the spin quantum numbers have been suppressed.
For the first (second) Cr ion ($\epsilon =\pm 1$), one replaces $z$ in the
RHS of (\ref{states}) by ($z \mp z_o$) respectively. The above
expression also includes the effect of the hemihedral part of the
trigonal distortion which is the most dominant interaction. This
interaction is of the form $\eta z$ and it leads to a mixing of
Cr $3d$ states and Cr $4p$ states with coefficients $\eta_1$, $\eta_2$
and $\eta_3$ being proportional to $\eta$.
Using (\ref{states}), together with the spin-orbit interaction,
one obtains after a lenghtly, but straightforward calculation,
the dynamical current operator
$ {\bf J}({\bf r},t)=\langle \Psi(t)|\hat{\bf J}|\Psi(t) \rangle$.
Here we report the results of this calculation, we will discuss
the details elsewhere \cite{bigpaper}.
The source term entering (\ref{S}) is then obtained to be
$$
{\bf S}({\bf r},t) = {4\pi\over c^2}
{\partial\over \partial t} {\bf J}({\bf r},t)~~.
$$
Since the full D$_{3d}$ symmetry of Cr$_2$O$_3$ is taken
into account, the calculations based on the (CrO$_6$)$_2$ model
correctly predict all selection rules also found by the macroscopic
approach. In addition, we also obtain estimates for all non-zero matrix
elements of the non-linear susceptibilities.

The interference between the MD and the ED processes in SHG can be observed
experimentally when the matrix elements, $\gamma_m$ and
$\chi_e$ occuring in (\ref{source}) are roughly of the same order of
magnitude.  We estimate their relative order of magnitude by
\begin{equation}
{\chi_e\over\gamma_m}\ \sim  4 \ {\lambda_0\over a_0}\
{\lambda\over E_e-E_{t_2}}\
{\eta\over E_p-E_d}\ \triangle(T)~~.
\label{estshg}
\end{equation}
Here $\lambda_0\approx 5000\AA$ is the wavelength of the emitted light,
$a_0\approx 0.69\AA$ is the radius of Cr$^{3+}$, $\lambda\approx 100$
cm$^{-1}$
is the spin-orbit interaction \cite{Rado,ST_58}, $E_{e}-E_{t_2}
\approx 8000 $cm$^{-1}$ is the difference in energy between the $t_2$ and
the $e$ orbitals, $\eta\approx 350$ cm$^{-1}$ is the trigonal field
\cite{ST_58}, $E_{p}-E_{d} \approx 8\times 10^4$ cm$^{-1}$ is the
difference in energy between the $d$ and the $p$ orbitals that are mixed
by the trigonal distortion and $\triangle(T) \approx 1$ is the
antiferromagnetic order parameter. The additional factor of 4 occurs
since there are 4 matrix elements of the same order of magnitude
(see (\ref{result})). The above expression gives the right
order of magnitude and we therefore conclude that the ED matrix element
in this mechanism can indeed interfere with the MD matrix element.
Clearly, the ED matrix element vanishes above $T_N$ when time reversal
symmetry is restored.

We now consider the phenomenon of gyrotropic birefringence (GB).  This is
another non-reciprocal effect, the possible existence of which was first
pointed out by Brown {\it et al} \cite{Brown}. GB is
a one-photon process appearing as a shift in the principal optic axes
along with a change in the velocity of propagation of light.
The first quantum
mechanical treatment of this problem was presented by Hornreich and
Shtrikman \cite{HS_68}, who estimated that GB in
Cr$_2$O$_3$ would lead to a shift in the optical axes of roughly
10$^{-8}$  rad, viz., a very small effect.  Recently however,
Krichevtsov {\it et al}. measured this non-reciprocal rotation and the
related magnetoelectric susceptibility of Cr$_2$O$_3$ in the optical
region \cite{Krichevtsov_93}. They found that the observed values were
at least 4 orders of magnitude larger than those predicted by
Hornreich and Shtrikman.
They also found that the temperature dependence of the non-reciprocal
effects mimicked that of the order parameter. The observed intensities
and temperature dependence suggest that these effects originate from
the ED process we have proposed.

Now it is known that the dominant contribution to GB is from the
magnetoelectric susceptibility defined by ${\bf M}^{(\omega)} = \alpha :
{\bf E}^{(\omega)}$ \cite{HS_68}.  Using (\ref{states}), we have calculated
the ED contribution to $\alpha$ in the optical region.
We find that
\begin{equation}
\alpha_{xx}  \sim ~ 4 \mu_o c e~{g\mu_B \over \hbar (\omega-\omega_n)}
{}~{\lambda\over E_e-E_{t_2}}\ {\eta\over E_p-E_d}~n_o~ \triangle(T)~,
\end{equation}
in dimensionless units. Here, $n_o$ is the density of Cr ions in
Cr$_2$O$_3$ ($\simeq 3.3 \times 10^{28}$ m$^{-3}$). In the region of
experimental interest, $\hbar(\omega-\omega_n) \sim 0.5$ eV.  Thus, we
estimate $\alpha_{xx} \sim 0.2 \times 10^{-4}$ which is of the same
order of magnitude as that observed experimentally. This also means that the
non-reciprocal rotation would be $\sim 10^{-4}$ rad. Since the ED
process we consider couples light to the order parameter, the observed
temperature dependence follows naturally from our mechanism.

To conclude, we have developed a microscopic model that explains all
non-reciprocal optical effects observed below $T_N$ in Cr$_2$O$_3$.
We have shown that these effects can be explained by an electric dipole
process that arises from an interplay between the spin-orbit coupling
and the trigonal distortion of the ligand field. Such a process
couples light directly to the antiferromagnetic order parameter.
Although we have applied the theory to explain non-reciprocal optical effects
in Cr$_2$O$_3$, it can be generalized to all materials where
i) the magnetic ion is not at a center of inversion and
ii) inversion symmetry is broken below $T_N$. In particular,
we predict that such effects should be observed in V$_2$O$_3$,
MnTiO$_3$ as also in the cuprate Gd$_2$CuO$_4$
below the ordering temperature of the Gadolinium magnetic subsystem,
$T_N$(Gd)=6.5K \cite{GdCuO}.
\acknowledgements
This work was stimulated by intensive discussions with M. Fiebig,
D. Fr\"{o}hlich, G. Sluyterman v. L. and H. J. Thiele.  This work was
supported by the Deutsche Forschungsgemeinschaft, the Graduiertenkolleg
``Festk\"{o}rperspektroskopie'' and by the European Community Human Capital
and Mobility program.

%
%
%
\begin{figure}
\caption{
         An illustration of the (CrO$_6$)$_2$ cluster model.
         The circles and the triangles indicate the positions
         of the Oxygen and Chromium ions respectively.
         The direction of the triangles indicate
         the ordering of the Cr$^{3+}$ moments in the antiferromagnetic state
         at a given domain.
         The cross is the location of the center of inversion.
\label{model}
             }
\end{figure}
\end{document}